\documentclass[a4paper, onecolumn, prX, showkeys]{revtex4}
\usepackage{graphicx}

\begin{document}

\title{A New Relativistic Orthogonal States Quantum Key Distribution Protocol}

\author{Jordan S. Cotler}
\affiliation{Massachusetts Institute of Technology, Cambridge, Massachusetts 02139, USA}
\author{Peter W. Shor}
\affiliation{Department of Mathematics, Center for Theoretical Physics and CSAIL, Massachusetts Institute of Technology, Cambridge, Massachusetts 02139, USA}

\begin{abstract}
We introduce a new relativistic orthogonal states quantum key distribution protocol which leverages the properties of both quantum mechanics and special relativity to securely encode multiple bits onto the spatio-temporal modes of a single photon.  If the protocol is implemented using a single photon source, it can have a key generation rate faster than the repetition rate of the source, enabling faster secure communication than is possible with existing protocols.  Further, we provide a proof that the protocol is secure and give a method of implementing the protocol using line-of-sight and fiber optic channels.
\end{abstract}


\maketitle

\section{Introduction}
Cryptography underpins all secure communications, whether it is used for transferring credit card information between a buyer and a seller through
the Internet, or relaying classified information over a military network. Classical cryptography is unsatisfactory in several respects.
Its security is generally based on assumptions about computation which are believed true but for which no absolute proof exists. That is, the assumptions depend on relative results which say that if certain computational problems are not efficiently solvable, then specific cryptosystems are secure.  In practical terms, classical cryptosystems frequently suffer from difficulties in secure key distribution and often need to be revised and updated to maintain their security in response to strong advances in the capabilities of modern computer systems.  In contrast, quantum cryptography was devised in the mid-1980's in order to create cryptographic protocols that are guaranteed to be secure by the laws of nature [1] (i.e. physics, and specifically quantum mechanics).

The BB84 protocol was the first practical quantum cryptography protocol to be proven secure using the laws of quantum mechanics [2, 3], and it has provided the framework for most implementations of quantum cryptography.  One major shortcoming of BB84 is its inefficient use of quantum mechanical information in that the key generation rate of a shared secret key is less than half of the repetition rate of the single photon source used for the protocol [1].  In 1995, Goldenberg and Vaidman proposed the first orthogonal states-based protocol which combined principles of special relativity with quantum mechanics to allow for secure communication [4].  Since then, variations of their protocol have been proposed [5, 6].  The way in which a relativistic quantum key distribution protocol exploits the laws of quantum mechanics is by creating a situation in which an eavesdropper must cause a detectable time delay if she wishes to obtain information about the secret key.  The eavesdropper cannot undo this delay without using faster-than-light communication - a task which is impossible according to special relativity.  In this paper we present a related but more efficient and flexible cryptographic protocol for which security is similarly guaranteed by nature and which expands the reach of provably secure cryptography.  Our QKD protocol has the unique feature that it can achieve a key generation rate greater than the repetition rate of the single photon source used for the protocol.

\section{The Protocol}
\subsection{Procedure}
According to common nomenclature, the sender of a cryptographic message is referred to as Alice, the receiver as Bob, and the eavesdropper as Eve.  The proposed protocol generates a secret key between Alice and Bob by encoding information on the phase between four states of a coherent superposition of single photons.  The state of each photon carries two bits of information.  The protocol is designed so that if Eve attempts to gain information about the secret key by manipulating the coherent superposition of any photon, either the photon will reach Bob detectably late, or the photon will be in a detectably different state of superposition.

In order to describe the protocol, we will follow the path of a single photon on its journey from Alice to Bob.  We will assume that communications are noiseless and lossless, meaning that no additional photons are introduced into the system and no photons are taken out of the system.  Further, we will assume that the channel between Alice and Bob is the shortest path between them (we will revisit this assumption later).  Let us also say that Alice and Bob share synchronized clocks.  Consider the setup in Figure 1.

\begin{center}
\includegraphics[scale=0.5]{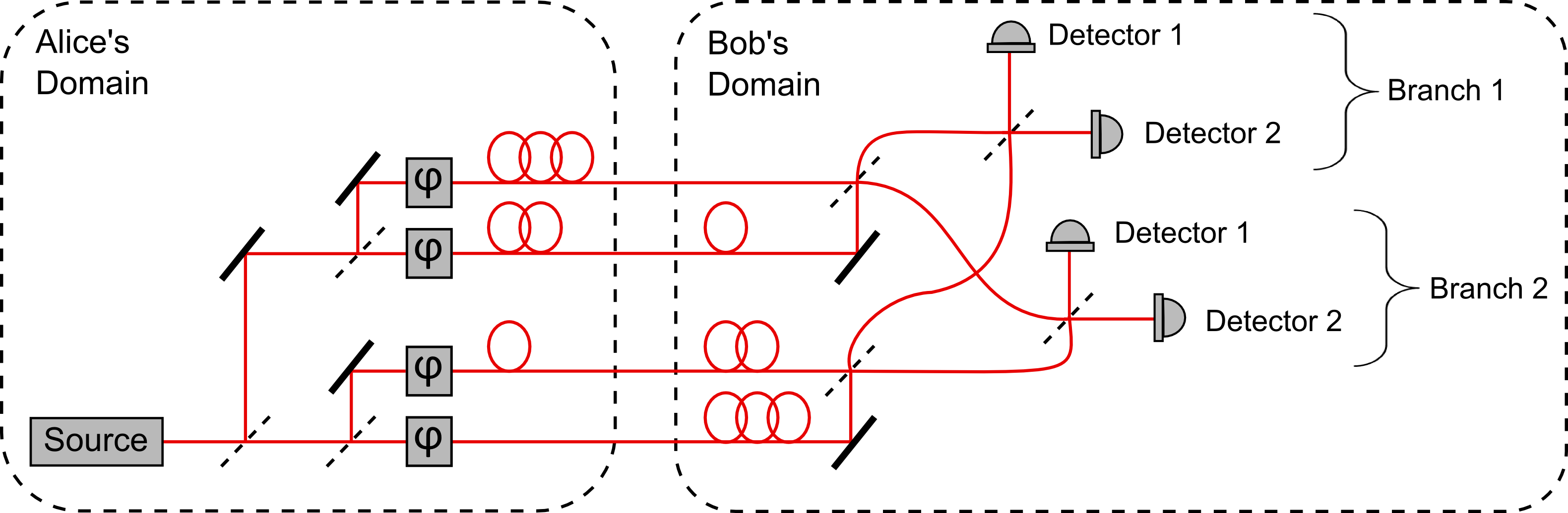}\\
${}$  \\
Fig. 1.  The setup of the protocol.
\end{center}

In Alice's domain, a single photon is emitted from Alice's single photon source.  Since single photon sources emit photons probabilistically, the photon is emitted at a random time.  Essential to the protocol is the window of time or ``time bin'' in which the photon will ultimately be received by Bob at one of his detectors.  The amount of time it takes for a photon to reach Bob's detectors depends upon how much the photon is delayed during its travels.  Next, the photon will pass through a 50-50 beam-splitter and will become a coherent superposition of a state on two separate paths.  Afterwards, the states on each of the paths pass through 50-50 beamsplitters, putting the photon into a coherent superposition of a state on four separate paths.  We will denote the component of the state on the bottom path by $|a\rangle$, the component on the second to bottom path by $|b\rangle$, the component on the second to top path by $|c\rangle$, and the component on the top path by $|d\rangle$.

The components of the photon are then phase-shifted randomly by Alice so that the photon is in one of the four states
\begin{equation}  |\Psi_0\rangle=\frac{1}{2}\left[
\left(|a\rangle+|b\rangle\right)+\left(|c\rangle+|d\rangle\right)
   \right] \end{equation}
\begin{equation}  |\Psi_1\rangle=\frac{1}{2}\left[
\left(|a\rangle-|b\rangle\right)+\left(|c\rangle-|d\rangle\right)
   \right] \end{equation}
\begin{equation}  |\Psi_2\rangle=\frac{1}{2}\left[
\left(|a\rangle+|b\rangle\right)-\left(|c\rangle+|d\rangle\right)
   \right] \end{equation}
\begin{equation}  |\Psi_3\rangle=\frac{1}{2}\left[
\left(|a\rangle-|b\rangle\right)-\left(|c\rangle-|d\rangle\right)
   \right] \end{equation}
Note that there are loops in Fig. 1 which symbolize additional lengths in the path of each component of the photon.  If the distance between Alice and Bob is $D$, we will take the length of each loop to be $D$.  We will later show that the length of each loop can be made very short.  If the length is equal to $D$, no two components of the state are accessible to Eve at the same time.

On Bob's receiving end of the setup, he adds length to the paths of the components of the state so that he can bring them all to one point in space and time.  Bob then interferes the $|a\rangle$ and $|b\rangle$ components with a 50-50 beamsplitter, and interferes the $|c\rangle$ and $|d\rangle$ components with a second 50-50 beamsplitter.  If Alice sent the state $|\Psi_1\rangle$ or $|\Psi_3\rangle$, the two interfered components of the state will end up on Branch 1.  Alternatively, if Alice sent the state $|\Psi_0\rangle$ or $|\Psi_2\rangle$, the two interfered components of the state will end up on Branch 2.  Whichever branch the components end up on, the two interfered components of the state are then interfered with one another by a final 50-50 beamsplitter.  Note that if Alice sent either the state $|\Psi_2\rangle$ or $|\Psi_3\rangle$, the photon will be received by one of the two detectors marked Detector 1, and if Alice sent either the state $|\Psi_0\rangle$ or $|\Psi_1\rangle$ the photon will be received by one of the two detectors marked Detector \nolinebreak 2.

Once Bob has received all of the states that Alice has sent, he broadcasts to Alice the time at which he received each state (i.e., the time that each photon reached one of his detectors).  Additionally, for each state that he received, Bob decides randomly whether to communicate on which branch he received the state (either Branch 1 or Branch 2) or at which detector he received the state (either Detector 1 or Detector 2).  Note that since Alice is randomly choosing the states which she sends to Bob, there is no correlation between the ``which-branch" information and the ``which-detector" information.  As a result, if for a particular state Bob broadcasts to Alice the which-branch information, they will use the undisclosed which-detector information as a bit for their shared secret key.  On the other hand, if for a particular state Bob broadcasts to Alice the which-detector information, they will use the undisclosed which-branch information as a bit for their shared secret key.  If Alice tells Bob that he either received a state at an incorrect time or that some which-branch or which-detector information was incorrect, then Bob knows that an eavesdropper must have been present.

In the protocol, Alice and Bob can achieve a key generation rate equal to the repetition rate of Alice's single photon source since two bits are encoded onto the spatio-temporal modes of a single photon, and only one of those bits is discarded for security. To achieve a key generation rate greater than the repetition rate of the source, the protocol may be modified by increasing the number of paths into which the photon is split.  In the modified protocol, Alice will put each photon in an equal superposition of a state on $2^k$ separate paths where one component at a time is sent to Bob's domain.  Alice can choose to add a relative phase of $\pi$ to combinations of components of the state, leading to $2^k$ states that she can send to Bob.  Once Bob has received all of the components of the photon, he interferes them such that the photon will end up at one of $2^k$ detectors, each one corresponding to one of the $2^k$ states that Alice can send.  It follows that in this modified protocol, the superposition state of each photon encodes $k$ bits, and some subset of those bits can be crosschecked in order to detect Eve.  Setting $k=2$, we recover the protocol which is the subject of this paper.

In practice, single photon sources are expensive and very difficult to fabricate while channels, such as fiber optic cables, are an abundant resource.  Therefore, it is most efficient for quantum key distribution protocols to optimize the number of secure bits that can be encoded per each emitted photon.  Our protocol provides a significant practical advantage over other existing protocols since it optimizes the number of bits that are encoded per each photon by using higher-dimensional channels.

\subsection{Security proof}
The security proof which follows is inspired by the structure of a proof given by Goldenberg and Vaidman for their own protocol [4]. We will prove that if Eve tries to gain any information about the secret key from any photon sent from Alice to Bob, then either Bob will receive the photon at the wrong time, or the photon will be in a detectably different state of superposition.  Eve cannot add or subtract photons from the system without inducing detectable noise or loss.  The only way that Eve could attempt to avoid detection is by preserving both the phase and the timing of a photon's wavefunction.  We will consider two times $t_1$ and $t_2$, the first of which is before any component of the photon has left Alice's domain, and the second of which is after all components of the photon have entered Bob's domain.  The most general operation that Eve can perform on the state is that of a superoperator.  In this case, we take $\textbf{U}$ to be one of the Kraus components of Eve's superoperator which takes a state at time $t_1$ to a state at time $t_2$.  For all four possible states that Alice can send, the free time-evolution with no eavesdropper is 
   
 \begin{equation}   |\Psi_0(t_1)\rangle \longrightarrow |\Psi_0(t_2)\rangle =\frac{1}{2}\left[
\left(|a(t_2)\rangle+|b(t_2)\rangle\right)+\left(|c(t_2)\rangle+|d(t_2)\rangle\right)
   \right] \end{equation}
\begin{equation} |\Psi_1(t_1)\rangle \longrightarrow |\Psi_1(t_2)\rangle = \frac{1}{2}\left[
\left(|a(t_2)\rangle-|b(t_2)\rangle\right)+\left(|c(t_2)\rangle-|d(t_2)\rangle\right)
   \right] \end{equation}
\begin{equation}|\Psi_2(t_1)\rangle \longrightarrow |\Psi_2(t_2)\rangle =\frac{1}{2}\left[
\left(|a(t_2)\rangle+|b(t_2)\rangle\right)-\left(|c(t_2)\rangle+|d(t_2)\rangle\right)
   \right] \end{equation}
\begin{equation} |\Psi_3(t_1)\rangle \longrightarrow |\Psi_3(t_2)\rangle = \frac{1}{2}\left[
\left(|a(t_2)\rangle-|b(t_2)\rangle\right)-\left(|c(t_2)\rangle-|d(t_2)\rangle\right)
   \right] \end{equation}

Since Eve does not know whether Bob will communicate to Alice the which-branch or which-detector information, she must prepare for either scenario.  In the case Bob tells Alice the which-branch information for a given state, Eve must make the absolute squares of the amplitudes of the $|a(t_2)\rangle$ and $|b(t_2)\rangle$ components have the same value, and the absolute squares of the amplitudes of the $|c(t_2)\rangle$ and $|d(t_2)\rangle$ components also have the same value.  Additionally, the relative phase between the $|a(t_2)\rangle$ and $|b(t_2)\rangle$ components as well as the relative phase between the $|c(t_2)\rangle$ and $|d(t_2)\rangle$ components must remain unchanged.  In the case Bob shares the which-detector information for a given state, Eve must make sure that the sum of the absolute squares of the amplitudes of the $|a(t_2)\rangle$ and $|b(t_2)\rangle$ components equals the sum of the absolute squares of the amplitudes of the $|c(t_2)\rangle$ and $|d(t_2)\rangle$ components.  Further, the relative phase between the $|a(t_2)\rangle$ and $|c(t_2)\rangle$ components as well as the relative phase between the $|b(t_2)\rangle$ and $|d(t_2)\rangle$ components must remain unchanged.  So as to account for both scenarios, it follows that Eve must make the magnitude of each of the final amplitudes equal to a constant $C$.  Eve must also preserve the relative phases between any two components of the state.  

Let $|\Phi(t)\rangle$ be the state of an auxiliary system which Eve uses to extract information.  It follows that for $i=0,1,2,3$, the general form of the evolution from time $t_1$ to $t_2$ is
      
   \begin{equation} \textbf{U}\,|\Psi_0(t_1)\rangle \, |\Phi(t_1)\rangle= 2C |\Psi_0(t_2)\rangle \, |\Phi_0(t_2)\rangle 
  \end{equation}
  \begin{equation} \textbf{U}\,|\Psi_1(t_1)\rangle \, |\Phi(t_1)\rangle= 2C |\Psi_1(t_2)\rangle \, |\Phi_1(t_2)\rangle 
  \end{equation}
    \begin{equation} \textbf{U}\,|\Psi_2(t_1)\rangle \, |\Phi(t_1)\rangle= 2C |\Psi_2(t_2)\rangle \, |\Phi_2(t_2)\rangle 
  \end{equation}
    \begin{equation} \textbf{U}\,|\Psi_3(t_1)\rangle \, |\Phi(t_1)\rangle= 2C |\Psi_3(t_2)\rangle \, |\Phi_3(t_2)\rangle 
  \end{equation}

\noindent where $|\Phi_i (t_2)\rangle$ is the state of Eve's auxiliary system at time $t_2$ after it has been acted upon by $\textbf{U}$ in the case that Alice sends the state $|\Psi_i \rangle$.  Note that if
  
 \begin{equation} |\Phi_0(t_2)\rangle=|\Phi_1(t_2)\rangle=|\Phi_2(t_2)\rangle=|\Phi_3(t_2)\rangle
 \end{equation}
 it is impossible for Eve to extract information.

Consider a photon at time $t_1$ in the state
 \begin{equation} |d(t_1)\rangle =\frac{1}{2}\left(|\Psi_0(t_1)\rangle -|\Psi_1(t_1)\rangle-|\Psi_2(t_1)\rangle+|\Psi_3(t_1)\rangle \right)
 \end{equation}
  It follows that $\textbf{U}\,|d(t_1)\rangle \, |\Phi(t_1)\rangle$ can be written as 
\begin{eqnarray}
 \textbf{U}\,|d(t_1)\rangle \, |\Phi(t_1)\rangle =C \,[                                     |a\rangle \,(|\Phi_0(t_2)\rangle-|\Phi_1(t_2)\rangle-|\Phi_2(t_2)\rangle+|\Phi_3(t_2)\rangle)\,\nonumber\\                              +|b\rangle \,(|\Phi_0(t_2)\rangle+|\Phi_1(t_2)\rangle-|\Phi_2(t_2)\rangle-|\Phi_3(t_2)\rangle)\,\,\nonumber\\                                   +|c\rangle \,(|\Phi_0(t_2)\rangle-|\Phi_1(t_2)\rangle+|\Phi_2(t_2)\rangle-|\Phi_3(t_2)\rangle)\,\,\nonumber\\                                +|d\rangle \,(|\Phi_0(t_2)\rangle+|\Phi_1(t_2)\rangle+|\Phi_2(t_2)\rangle+|\Phi_3(t_2)\rangle)]
\end{eqnarray}
  This equation expresses that unless
 
  \begin{eqnarray}
         |\Phi_0(t_2)\rangle-|\Phi_1(t_2)\rangle-|\Phi_2(t_2)\rangle+|\Phi_3(t_2)\rangle=0 \nonumber \\ \nonumber \\
         |\Phi_0(t_2)\rangle+|\Phi_1(t_2)\rangle-|\Phi_2(t_2)\rangle-|\Phi_3(t_2)\rangle=0 \nonumber\\ \nonumber\\
         |\Phi_0(t_2)\rangle-|\Phi_1(t_2)\rangle+|\Phi_2(t_2)\rangle-|\Phi_3(t_2)\rangle=0
\end{eqnarray}
the photon could be measured in a state $|a(t_2)\rangle$, $|b(t_2)\rangle$, or $|c(t_2)\rangle$.  However, the photon cannot be measured to be in any of these three states.  The reason is that $\textbf{U}$ is an operator applied by Eve that takes a photon in a state $|d(t_1)\rangle$ to a coherent superposition of the states $|a(t_2)\rangle$, $|b(t_2)\rangle$, $|c(t_2)\rangle$ and $|d(t_2)\rangle$.  But a photon in a state  $|a(t_2)\rangle$, $|b(t_2)\rangle$, or $|c(t_2)\rangle$ must have exited Eve's domain before the component in the state $|d(t_1)\rangle$ has entered it.  Thus, for Eve to preserve causality, the coefficients of $|a\rangle$, $|b\rangle$ and $|c\rangle$ in Eq. (15) must be zero which immediately leads to Eq. (16), or equivalently
$$|\Phi_0(t_2)\rangle=|\Phi_1(t_2)\rangle=|\Phi_2(t_2)\rangle=|\Phi_3(t_2)\rangle$$
and so Eve cannot extract any information.  This completes the proof.  Note that a security proof of the modified protocol presented at the end of Section II.A is a straightforward extension of the above proof.

\section{Implementation and Variations}
\subsection{Shortening the Loop Length}
In the above explanation of the protocol, we assumed that the length of each loop in the setup was $D$ where $D$ is the distance between Alice and Bob.  In practice, having such long loops would decrease the key generation rate dramatically, and would also lead to photon loss due to the large distances that each component of the photon would have to travel.  The solution to this problem was originally proposed by Goldenberg and Vaidman in the context of their own protocol [4].  Let us say that the accuracy of the time of Alice and Bob's measurements is $\Delta t$.  Therefore, if Bob receives a photon that has been delayed by any amount of time greater than or equal to $\Delta t$ relative to Alice's sending time, he and Alice will be able to detect this delay.

Let us now say that the length of each loop is $c\,(\Delta t+\epsilon)$ for some positive $\epsilon$.  It is easy to see that the terms $|a(t_2)\rangle$, $|b(t_2)\rangle$, or $|c(t_2)\rangle$ still cannot appear in $\textbf{U}\,|d(t_1)\rangle \, |\Phi(t_1)\rangle$.  Therefore, the security proof will still hold for loop lengths of $c\,(\Delta t+\epsilon)$, and it is practical to incorporate this change into the protocol.

\subsection{Mode Coupling and Sub-optimal Paths}
In implementing the protocol experimentally, it is not practical to have components of a photon travel along four separate paths over the large distances that may separate the domains of Alice and Bob.  Optical fibers tend to expand and contract with small temperature changes, which can change the relative phase between the components of the photon thereby affecting how they will eventually interfere with one another in Bob's domain.  Atmospheric turbulence provides similar difficulties for free-space transmission of photons.  Both problems can seriously degrade the effectiveness of the protocol, and potentially reduce the secret key generation rate.

The problem can be solved by using switches to couple the four paths to four temporal modes of one fiber or one free-space transmission line.  The photon then travels in a single spatial mode from Alice's domain to Bob's domain.  By analogy, imagine four trains traveling on four separate tracks.  By using track switches to bring the trains onto a single track, one can align the trains so that they follow one another.  The process can later be reversed, allowing the trains to be switched back onto four separate tracks.  Similarly, the four components of the photon traveling along the four paths are coupled to one spatial mode using switches S$1$, S$2$, and S$3$ in Alice's domain (Figure 2).  The photons then travel one after the other in a single fiber or free-space path until reaching Bob's domain where switches S$4$, S$5$, and S$6$ couple their temporal modes back into four spatial modes.

\begin{center}
\includegraphics[scale=0.5]{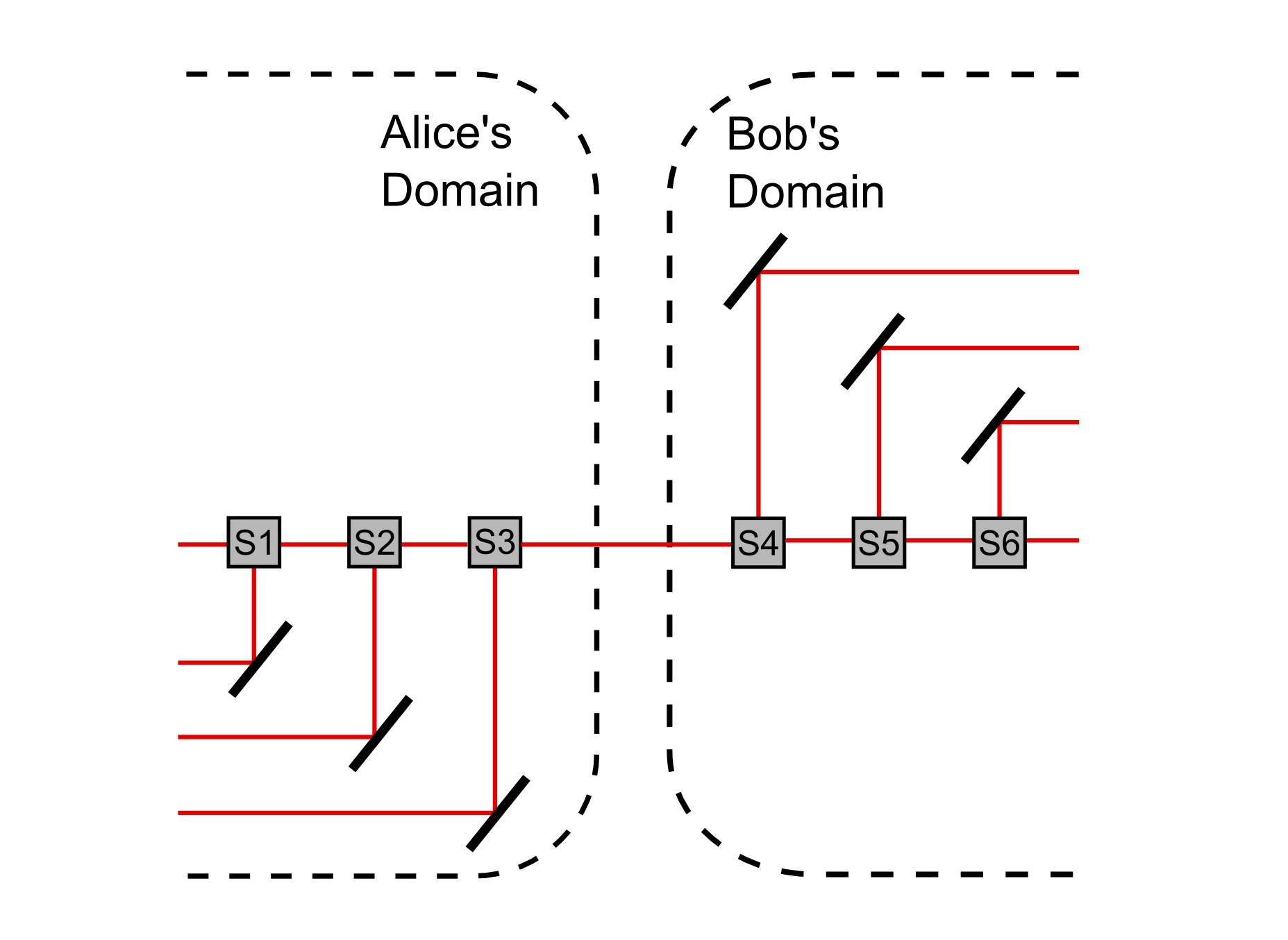}\\
${}$  \\
Fig. 2.  Coupling spatial modes to temporal modes.
\end{center}

Recall that the protocol assumes that the path of the photons between the domains of Alice and Bob is a straight line.  In satellite communication, this assumption is correct since the communication is line-of-sight.  However, in fiber optic communications, fibers are never laid down in straight lines.  If the path of the photons is not a straight line, an eavesdropper could create a shorter straight-line path between Alice and Bob thereby reducing the travel time of the photons.  As a result, the eavesdropper can allow herself extra time in which to measure and delay Alice's photons and still send them on to Bob such that he will receive the photons at the expected time. Therefore, using a suboptimal path could mask the presence of an eavesdropper.

The protocol can be modified to address the problem of a suboptimal path.  For such a modification, Alice and Bob must take into account the difference in length between their suboptimal path and an optimal straight-line path between them.  The difference in distance between the two paths will be designated as $\mu$.  To make the protocol secure and account for this extra distance, each loop in the setup must be increased in length by at least $\mu$.  However, this change need not decrease the efficiency of the protocol since a setup in which the temporal modes are coupled to the spatial mode of a single fiber could be designed such that components of one photon are spatially interlaced with the components of other photons.  It is important to note that Alice and Bob's switches and setup of interferometers have to be designed to function in accordance with the way that the photon components are interlaced.

\section{Conclusion}

In our information age, large amounts of data need to be transmitted securely and rapidly for commercial and military purposes.  The increasing power of computers makes conventional non-quantum cryptography protocols susceptible to attack. Quantum cryptography offers benefits over standard methods of cryptography because it provides fundamental security. The protocol presented in this paper is a new method for relativistic orthogonal states quantum key distribution which provides security and achieves a key generation rate higher than other known QKD protocols. The efficiency of the proposed protocol makes it an attractive alternative to existing QKD protocols for both secure fiber optic as well as secure satellite communication.

\section*{Acknowledgements}

This research was funded by the Undergraduate Research Opportunities Program at the Massachusetts Institute of Technology.  The authors wish to thank Joseph Altepeter and Steven Homer for their useful discussions and support.

\end{document}